\documentclass[letterpaper]{sig-alternate-2013}
\usepackage{booktabs} 

\usepackage{graphicx}
\usepackage{epstopdf}
\usepackage{array}
\usepackage{pbox}
\usepackage{wrapfig}
\usepackage{adjustbox}

\usepackage{amsmath,amssymb,amsfonts}
\usepackage{mathtools}

\usepackage{url}
\usepackage{verbatim}
\usepackage{multirow}
\usepackage{rotating}
\usepackage{todonotes}

\usepackage[linesnumbered,ruled,vlined]{algorithm2e}
\usepackage{algpseudocode}

\usepackage{ifpdf}
\usepackage{lmodern}
\usepackage{lipsum}
\usepackage[T1]{fontenc}
\usepackage{textcomp}

\usepackage{enumitem}
\usepackage{mdwlist}

\usepackage{listings}

\usepackage{comment}
\usepackage{ifthen}
\usepackage{color}
\usepackage{colortbl,xcolor}

\usepackage{hhline}

\definecolor{LightCyan}{rgb}{0.88,1,1}
\definecolor{Gray}{gray}{0.9}
\definecolor{lightgray}{rgb}{0.83, 0.83, 0.83}
\definecolor{darkgray}{rgb}{0.66, 0.66, 0.66}
\specialcomment{notes}{\begingroup \color{blue}} { \endgroup}

\colorlet{punct}{red!60!black}
\definecolor{background}{HTML}{EEEEEE}
\definecolor{delim}{RGB}{20,105,176}
\colorlet{numb}{magenta!60!black}

\makeatletter
\newcommand*{\inlineequation}[2][]{%
  \begingroup
    \refstepcounter{equation}%
    \ifx\\#1\\%
    \else
      \label{#1}%
    \fi
    \relpenalty=10000 %
    \binoppenalty=10000 %
    \ensuremath{%
      #2%
    }%
    ~\@eqnnum
  \endgroup
}
\makeatother

\lstdefinelanguage{json}{
    basicstyle=\small\ttfamily,
    numbers=left,
    numbers=none,
    stepnumber=1,
    numbersep=8pt,
    showstringspaces=false,
    breaklines=true,
    frame=lines,
    backgroundcolor=\color{background},
    literate=
     *{0}{{{\color{numb}0}}}{1}
      {1}{{{\color{numb}1}}}{1}
      {2}{{{\color{numb}2}}}{1}
      {3}{{{\color{numb}3}}}{1}
      {4}{{{\color{numb}4}}}{1}
      {5}{{{\color{numb}5}}}{1}
      {6}{{{\color{numb}6}}}{1}
      {7}{{{\color{numb}7}}}{1}
      {8}{{{\color{numb}8}}}{1}
      {9}{{{\color{numb}9}}}{1}
      {:}{{{\color{punct}{:}}}}{1}
      {,}{{{\color{punct}{,}}}}{1}
      {\{}{{{\color{delim}{\{}}}}{1}
      {\}}{{{\color{delim}{\}}}}}{1}
      {[}{{{\color{delim}{[}}}}{1}
      {]}{{{\color{delim}{]}}}}{1},
}

\newcolumntype{!}{>{\global\let\currentrowstyle\relax}}
\newcolumntype{^}{>{\currentrowstyle}}

\DeclareGraphicsExtensions{.eps,.pdf,.png,.jpeg}
\graphicspath{{figure/eps/},{figure/pdf/},{figure/png/},{figure/jpeg/}}

\newcommand{\superscript}[1]{\ensuremath{^{\textrm{#1}}}}

\newcommand{\si}{\begin{enumerate}[leftmargin=*, itemindent=0cm, align=left]\itemsep0em}

\newcommand{\ei}{\end{enumerate}}

\usepackage{subcaption}
\usepackage[utf8]{inputenc}
\lstdefinelanguage{json}{
    basicstyle=\normalfont\ttfamily,
    numbers=left,
    numberstyle=\scriptsize,
    stepnumber=1,
    numbersep=8pt,
    showstringspaces=false,
    breaklines=true,
    frame=lines,
    tabsize=2,
    backgroundcolor=\color{background},
    literate=
     *{0}{{{\color{numb}0}}}{1}
      {1}{{{\color{numb}1}}}{1}
      {2}{{{\color{numb}2}}}{1}
      {3}{{{\color{numb}3}}}{1}
      {4}{{{\color{numb}4}}}{1}
      {5}{{{\color{numb}5}}}{1}
      {6}{{{\color{numb}6}}}{1}
      {7}{{{\color{numb}7}}}{1}
      {8}{{{\color{numb}8}}}{1}
      {9}{{{\color{numb}9}}}{1}
      {:}{{{\color{punct}{:}}}}{1}
      {,}{{{\color{punct}{,}}}}{1}
      {\{}{{{\color{delim}{\{}}}}{1}
      {\}}{{{\color{delim}{\}}}}}{1}
      {[}{{{\color{delim}{[}}}}{1}
      {]}{{{\color{delim}{]}}}}{1},
}

\usepackage{tikz}
\usepackage{longtable}

\usetikzlibrary{arrows}

\definecolor{Gray}{gray}{0.9}
\definecolor{LightCyan}{rgb}{0.88,1,1}

\specialcomment{notes}{\begingroup \color{blue}} { \endgroup}

\makeatletter
\let\oldfootnote\footnote
\def\footnote{\@ifstar\footnote@star\footnote@nostar}
\def\footnote@star#1{{\let\thefootnote\relax\footnotetext{#1}}}
\def\footnote@nostar{\oldfootnote}
\makeatother

\DeclareGraphicsExtensions{.eps,.png}
\graphicspath{{figure/}{figure/eps/}{figure/png/}}

\SetKwInput{KwInput}{Input}
\SetKwInput{KwOutput}{Output}
\SetKwProg{myproc}{Procedure}{}{}
\DeclareMathOperator*{\argmax}{arg\,max}

  \renewenvironment{thebibliography}[1]{%
    \begin{oldthebibliography}{#1}%
      \setlength{\parskip}{0ex}%
      \setlength{\itemsep}{0ex}%
  }%
  {%
    \end{oldthebibliography}%
  }
  
\begin{document}

\title{High-Throughput and Language-Agnostic Entity Disambiguation and Linking on User Generated Data}

\def\kloutinc{\superscript{*}}
\def\rutgers{\superscript{\dag}}



\numberofauthors{1} 
\author{
    \alignauthor Preeti Bhargava, Nemanja Spasojevic, Guoning Hu \\
    \affaddr{Lithium Technologies | Klout}\\
    \affaddr{San Francisco, CA}\\
    \email{\{preeti.bhargava, nemanja.spasojevic, guoning.hu\}@lithium.com}
}

\maketitle

\begin{abstract}
The Entity Disambiguation and Linking (EDL) task matches \emph{entity mentions} in text to a unique Knowledge Base (KB) identifier such as a Wikipedia or Freebase id. It plays a critical role in the construction of a high quality information network, and can be further leveraged for a variety of information retrieval and NLP tasks such as text categorization and document tagging. EDL is a complex and challenging problem due to ambiguity of the mentions and real world text being multi-lingual. Moreover, EDL systems need to have high throughput and should be lightweight in order to scale to large datasets and run on off-the-shelf machines. More importantly, these systems need to be able to extract and disambiguate dense annotations from the data in order to enable an Information Retrieval or Extraction task running on the data to be more efficient and accurate. In order to address all these challenges, we present the Lithium EDL system and algorithm - a high-throughput, lightweight, language-agnostic EDL system that extracts and correctly disambiguates 75\% more entities than state-of-the-art EDL systems and is significantly faster than them.

\end{abstract}


\section{Introduction}
\label{section:introduction}
In Natural Language Processing (NLP), Entity Disambiguation and Linking (EDL) is the task of matching \emph{entity mentions} in text to a unique Knowledge Base (KB) identifier such as a Wikipedia or a Freebase id. It differs from the conventional task of Named Entity Recognition, which is focused on identifying the occurrence of an entity and its type but not the specific unique entity that the mention refers to. EDL plays a critical role in the construction of a high quality information network such as the Web of Linked Data \cite{heath2011linked}. Moreover, when any new piece of information is extracted from text, it is necessary to know which real world entity this piece refers to. If the system makes an error here, it loses this piece of information and introduces noise.

EDL can be leveraged for a variety of information retrieval and NLP tasks such as text categorization and document tagging. For instance, any document which contains entities such as \emph{Michael Jordan} and \emph{NBA} can be tagged with categories \emph{Sports} and \emph{Basketball}. It can also play a significant role in recommender systems which can personalize content for users based on the entities they are interested in.

EDL is complex and challenging due to several reasons:
\begin{itemize}[noitemsep,nolistsep]
\item Ambiguity - The same entity mention can refer to different real world entities in different contexts. A clear example of ambiguity is the mention \emph{Michael Jordan} which can refer to the basketball player in certain context or the machine learning professor from Berkeley. To the discerning human eye, it may be easy to identify the correct entity, but any EDL system attempting to do so needs to rely on contextual information when faced with ambiguity.
\item Multi-lingual content - The emergence of the web and social media poses an additional challenge to NLP practitioners because the user generated content on them is often multi-lingual. Hence, any EDL system processing real world data on the web, such as user generated content from social media and networks, should be able to support multiple languages in order to be practical and applicable. Unfortunately, this is a challenge that has not been given enough attention. 
\item High throughput and lightweight - State-of-the-art EDL systems should be able to work on large scale datasets, often involving millions of documents with several thousand of entities. Moreover, these systems need to have low resource consumption in order to scale to larger datasets in a finite amount of time. In addition, in order to be applicable and practical, they should be able to run on off-the-shelf commodity machines.
\item Rich annotated information - All information retrieval and extraction tasks are more efficient and accurate if the underlying data is rich and dense. Hence, EDL systems need to ensure that they extract and annotate many more entities and of different types (such as professional titles, sports, activities etc.) in addition to just named entities (such as persons, organizations, locations etc.) However, most existing systems focus on extracting named entities only.
\end{itemize}
In this paper, we present our EDL system and algorithm, hereby referred to as the Lithium EDL system, which is a high-throughput, lightweight  and language-agnostic EDL system that 
extracts and correctly disambiguates 75\% more entities than state-of-the-art EDL systems and is significantly faster than them.

\subsection{Related Work}

EDL has been a well studied problem in literature and has gained a lot of attention in recent years. Approaches that disambiguate entity mentions with respect to Wikipedia date back to Bunescu and Pasca's work in \cite{bunescu2006using}. 
Cucerzan \cite{cucerzan2007large} attempted to solve the same problem by using heuristic rules and Wikipedia disambiguation markups to derive mappings from display names of entities to their Wikipedia entries. However, this approach doesn't work when the entity is not well defined in their KB. Milne and Witten \cite{Milne:2008:LLW:1458082.1458150} refined Cucerzan's work by defining topical coherence using normalized Google Distance \cite{cilibrasi2007google} and only using `unambiguous entities' to calculate topical coherence. 

Recent approaches have focused on exploiting statistical text features such as mention and entity counts, entity popularity and context similarity to disambiguate entities. Spotlight \cite{daiber2013improving} used a maximum likelihood estimation approach using mention and entity counts. To combine different types of disambiguation knowledge together, Han and Sun \cite{han2011generative} proposed a generative model to include evidences from entity popularity, mention-entity association and context similarity in a holistic way. More recently, systems like AIDA \cite{yosef2011aida} and AIDA-light \cite{nguyen2014aida} have proposed graphical approaches that employ these statistical measures and attempt the disambiguation of multiple entries in a document simultaneously. Bradesco et al. \cite{bradesko2015isaac} followed an approach similar to AIDA-light \cite{nguyen2014aida} but limited the entities of interest to people and companies. However, a major disadvantage of such approaches is that their combinatorial nature results in intractability, which makes them harder to scale to very large datasets in a finite amount of time. In addition, all these systems do not support multi-lingual content which is very common nowadays due to the prolificity of user generated content on the web. 

Our work differs from the existing work in several ways. We discuss these in the contributions outlined below.

\subsection{Contributions}
Our contributions in this paper are:
\begin{itemize}[noitemsep,nolistsep]
\item Our EDL algorithm uses several context-dependent and context-independent features, such as mention-entity cooccurrence, entity-entity cooccurrence, entity importance etc., to disambiguate mentions to their respective entities.
\item In contrast to several existing systems such as Google Cloud NL API \footnote{\url{https://cloud.google.com/natural-language/}}, OpenCalais \footnote{\url{http://www.opencalais.com/}} and AIDA \cite{yosef2011aida}, our EDL system recognizes several types of entities (professional titles, sports, activities etc.) in addition to named entities (people, places, organizations etc.). Our experiments (Section \ref{sec:entitydensitycomparison}) demonstrate that it recognizes and correctly disambiguates about 75\% more entities than state-of-the-art systems. Such richer and denser annotations are particularly useful in understanding the user generated content on social media to model user conversations and interests.
\item Our EDL algorithm is language-agnostic and currently supports 6 different languages including English, Arabic, Spanish, French, German, and Japanese\footnote{Our EDL system can easily support more languages with the ready availability of ground truth data in them}. As a result, it is highly applicable to process real world text such as multi-lingual user generated content from social media. Moreover, it does not need any added customizations to support additional languages. In contrast, systems such as AIDA \cite{yosef2011aida} and AIDA-light \cite{nguyen2014aida} need to be extended by additional components in order to support other languages such as Arabic \cite{yosef2014aidarabic}.
\item Our EDL system has high throughput and is very lightweight. It can be run on an off-the-shelf commodity machine and scales easily to large datasets. Experiments with a dataset of 910 million documents showed that our EDL system took about 2.2ms per document (with an average size of 169 bytes) on a 2.5 GHz Xeon processor (Section \ref{sec:results}). Moreover, our experiments demonstrate that our system's runtime per unique entity extracted is about 3.5 times faster than state-of-the-art systems such as AIDA \cite{yosef2011aida}.
\end{itemize}

\section{Knowledge Base}
\label{section:knowledgebase}
Our KB consists of about 1 million Freebase\footnote{Freebase was a standard community generated KB until June 2015 when Google deprecated it in favor of the commercially available Knowledge Graph API.} machine ids for entities.  
These were chosen from a subset of  all Freebase entities that map to Wikipedia entities.
We  prefer to use Freebase rather than Wikipedia as our KB since in Freebase, the same id represents a unique entity across multiple languages.
Due to limited resources and usefulness of the entities, our KB contains approximately 1 million most important entities from among all the Freebase entities. 
This gives us a good balance between coverage and relevance of entities for processing common social media text.
Section \ref{subsection:OfflineDictionary} explains how entity importance is calculated, which enables us to rank the top 1 million Freebase entities.

In addition to the KB entities, we also employ two special entities: \textbf{NIL} and \textbf{MISC}.
\textbf{NIL} entity indicates that there is no entity associated with the mention, eg. mention `the' within the sentence may link to entity \textbf{NIL}. This entity is useful especially when it comes to dealing with stop words and false positives. \textbf{MISC} indicates that the mention links to an entity which is outside the selected entity set in our KB. 

\section{System Architecture}
\label{section:systemarch}

\begin{figure*}[t]
\centering
\includegraphics[width=\textwidth, height=15mm]{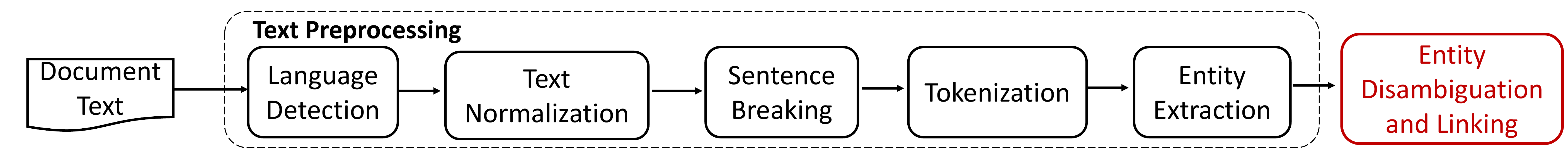}
\caption{Overview of the Lithium NLP pipeline}
\label{fig:papyrus}
\end{figure*}

This paper is focused on describing the Lithium EDL system. However, the EDL system is a component of a larger Natural Language Processing (NLP) pipeline, hereby referred to as the Lithium NLP pipeline, which we describe briefly here.
Figure \ref{fig:papyrus} shows the high level overview of the Lithium NLP pipeline. It consists of several Text Preprocessing stages before EDL.

\subsection{Text Preprocessing}

The Lithium NLP pipeline processes an input text document in the following stages before EDL:
\begin{itemize}[noitemsep,nolistsep]
\item \textbf{Language Detection} - This stage detects the language of the input document using a naive Bayesian filter. It has a 
precision of ~ $99\%$ and is available on GitHub\footnote{\url{https://github.com/shuyo/language-detection}}.
\item \textbf{Text Normalization} - This stage normalizes the text by escaping unescaped characters and replacing some special characters based on the detected language. For example, it replaces non-ASCII punctuations with spaces and converts accents to regular characters for English. 
\item \textbf{Sentence Breaking} - This stage breaks the normalized text into sentences using the Java Text API\footnote{\url{https://docs.oracle.com/javase/7/docs/api/java/text/BreakIterator.html}}.
This tool can distinguish sentence breakers from other marks, such as periods within numbers and abbreviations, according to the detected language.
\item \textbf{Tokenization} - This stage converts each sentence into a sequence of tokens via the 
Lucene Standard Tokenizer\footnote{\url{http://lucene.apache.org/core/4_5_0/analyzers-common/org/apache/lucene/analysis/standard/StandardTokenizer.html}}.
\item \textbf{Entity Extraction} -  This stage captures mentions in each sentence that belong to precomputed offline dictionaries.  
Please see Section \ref{subsection:OfflineDictionary} for more details about dictionary generation. 
A mention may contain a single token or several consecutive tokens, but a token can belong to at most one mention.
Often there are multiple ways to break a sentence into a set of mentions.
To make this task computationally efficient, we apply a simple greedy strategy that analyzes windows of \emph{n}-grams (n $\in$ [1,6]) and extracts the longest mention found in each window.
\end{itemize}

An extracted mention maps to multiple candidate entities.
Our pipeline determines the best entity for each mention in the EDL phrase, which is described in Section \ref{subsec:EDL}.


\subsection{Data Set Generation\label{data_set}}

Since our goal here is to build a language-agnostic EDL system, we needed a dataset that scales across several languages and also has good entity density and coverage. Unfortunately, such a dataset is not readily available. 
Hence, we generated a ground truth data set for our EDL system, the Densely Annotated Wikipedia Text (DAWT)\footnote{DAWT and other derived datasets are available for download at: \url{https://github.com/ klout/opendata/tree/master/wiki_annotation}.} \cite{Spasojevic:dawt}, 
using densely Wikified \cite{Mihalcea:2007:WLD:1321440.1321475} or annotated Wikipedia articles. 
Wikification is entity linking with Wikipedia as the KB. We started with Wikipedia data dumps\footnote{\url{https://dumps.wikimedia.org/}},
which were further enriched by introducing more hyperlinks in the existing document structure. 
Our main goals when building this data set were to maintain high precision and increase linking coverage. 
As a last step, the hyperlinks to Wikipedia articles in a specific language were replaced with links to their Freebase ids to adapt to our KB. The densely annotated Wikipedia articles had on an average 4.8 times more links than the original articles.

\begin{figure}[t]
\centering
\includegraphics[width=0.48\textwidth,height=55mm]{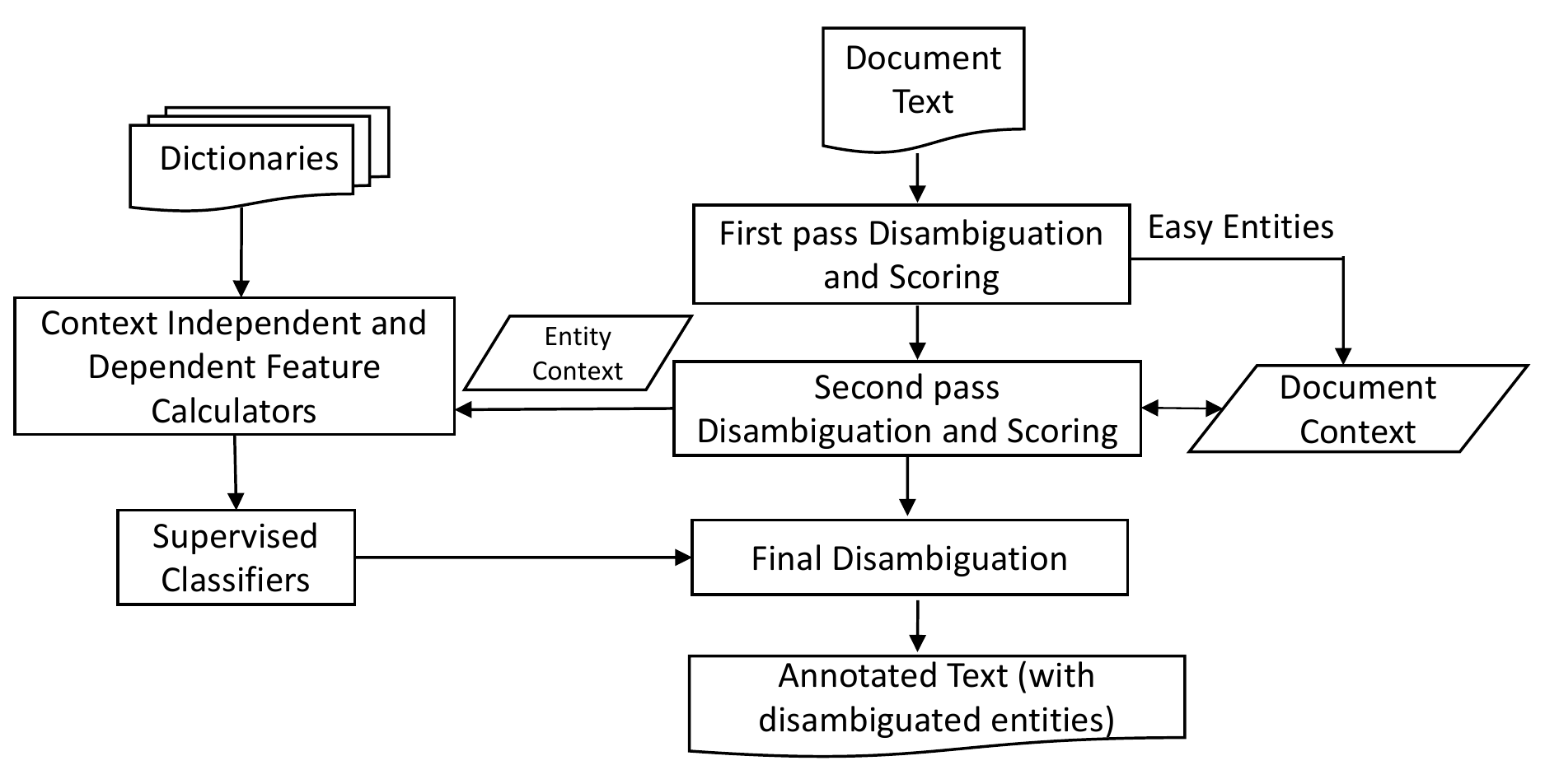}
\caption{System architecture of the Entity Disambiguation and Linking stage}
\label{fig:systemarch}
\end{figure}

\subsection{Entity Disambiguation and Linking}\label{subsec:EDL}

The system architecture of the EDL stage is shown in Figure \ref{fig:systemarch}. Similar to the approach employed by AIDA-light \cite{nguyen2014aida}, it employs a two-pass algorithm (explained in detail in Section \ref{section:algorithm}) which first identifies a set of \emph{easy} mentions, which have low ambiguity and can be disambiguated and linked to their respective entities with high confidence. It then leverages these easy entities and several context dependent and independent features to disambiguate and link the remaining \emph{hard} mentions. However, unlike AIDA-light \cite{nguyen2014aida}, our approach does not use a graph based model to jointly disambiguate entities because such approaches can become intractable with increase in the size of the document and number of entities. In addition, our EDL problem is posed as a classification rather than a regression problem as in AIDA-light \cite{nguyen2014aida}.

The EDL stage consists of the following components:

\subsubsection{\textbf{Offline Dictionaries Generation}}
\label{subsection:OfflineDictionary}
Our EDL system uses several dictionaries capturing language models, probabilities and relations across entities and topics. 
These are generated by offline processes leveraging various multi-lingual data sources to generate resource files. These are:

\begin{itemize}[noitemsep,nolistsep]
  \item \textbf{Mention-Entity Cooccurrence} - This dictionary is derived using the DAWT data set \cite{Spasojevic:dawt}. Here, we estimate the prior probability that a mention $M_{i}$ refers to an entity $E_{j}$ (including \textbf{NIL} and \textbf{MISC}) with respect to our KB and corpora. It is equivalent to the cooccurrence probability of the mention and the entity:\small
$${{count (M_{i} \rightarrow E_{j})} \over {count (M_{i})}}$$
\normalsize
We generate a separate dictionary for each language. Moreover, since DAWT is 4.8 times denser than Wikipedia, these dictionaries capture several more mentions and are designed to be exhaustive across several domains. 
  \item \textbf{Entity-Entity Cooccurrence} - This dictionary is also derived using DAWT. In this case, we capture co-occurrence frequencies among entities by counting all the entities that simultaneously appear within a sliding window of 50 tokens. Moreover, this data is accumulated across all languages and is language independent in order to capture better relations and create a smaller memory footprint when supporting additional languages. Also, for each entity, we consider only the top 30 co-occurring entities which have at least 10 or more co-occurrences across all supported languages.
  \item \textbf{Entity Importance} - The entity importance score \cite{Bhattacharyya-importance} is derived as a global score identifying how important an extracted entity is for a casual observer. This score is calculated using linear regression with features capturing popularity within Wikipedia links, and importance of the entity within Freebase. We used signals such as Wiki page rank, Wiki and Freebase incoming and outgoing links, and type descriptors within knowledge base etc.

  \item \textbf{Topic Parent} - The Klout Topic Ontology\footnote{\url{https://github.com/klout/opendata}} is a manually curated ontology built to capture social media users' interests \cite{nemanja-lasta} and expertise scores \cite{Spasojevic2016:experts} across multiple social networks. As of December 2016, it consists of roughly 7,500 topic nodes and 13,500 edges encoding hierarchical relationships among them. The Topic Parents dictionary contains the parent topics for each topic within this ontology.

  \item \textbf{Entity To Topic Mapping} - 
  This dictionary essentially contains topics from the Klout Topic Ontology that are associated with the different entities in our KB. E.g. Michael Jordan, the basketball player, will be associated with the topics `Football' and `Sports'. We generate this dictionary via a weighted ensemble of several algorithms that employ entity co-occurrence and propagate the topic labels.
  A complete description of these algorithms is beyond the scope of this paper.

\end{itemize}

\subsubsection{\textbf{Context}}

\begin{itemize} [noitemsep,nolistsep]
\item \textbf{Document context} - As mentioned earlier, the Lithium EDL system relies on disambiguating a set of \emph{easy} 
mentions in the document which are then leveraged to disambiguate the \emph{hard} 
mentions. Thus, for each document, we maintain a \emph{document context} $C(T_{i})$ which includes all the easy entities in the document text that have been disambiguated. This context also includes cached pairwise feature scores for the context dependent features between the easy and hard entities (see Section \ref{sec:features} for a description of the context dependent features). 

\item \textbf{Entity context} - For each candidate entity $E_{k}$ of a hard mention, we define an \emph{entity context} $C'(E_{k})$ which includes the position of the corresponding mention in the document, the index number of the candidate entity as well as an \emph{easy entity window} $\mathbb{E}_{k}$ surrounding the hard mention. The appropriate window size \emph{W} is determined by parameter tuning on a validation set. 

\end{itemize}

\subsubsection{\textbf{Supervised Classifiers}}\label{subsubsec:class}

We pose our EDL problem as a binary classification problem for the following reason: For each mention, only one of the candidate entities is the correct label entity. Our ground truth data set provides the labeled correct entity but does not have any scores or ranked order for the candidate entities. Hence, we pose this problem as predicting one of the two labels \emph{$\left\{\text{True, False}\right\}$} for each candidate entity (where \emph{True} indicates it is the correctly disambiguated entity for a mention and \emph{False} indicates that it is not).

Using the process described in Section \ref{data_set}, we generated a ground truth training set of 70 English Wikipedia pages which had a total of 43,662 mentions and 147,236 candidate entities. We experimented with several classifiers such as Decision Trees, Random Forest, k-Nearest Neighbors and Logistic Regression on this training set. Decision Trees and Logistic Regression outperformed most of the classifiers. While Random Forest was as accurate as the Decision Tree classifier, it was computationally more expensive. Hence, we use Decision Tree and Logistic Regression in the Lithium EDL system.

\section{Entity Disambiguation and Linking Algorithm}
\label{section:algorithm}
\begin{algorithm*}[t]
    \SetAlgoLined
\KwInput{Text $T_{i}$ with extracted mentions $\mathbb{M}_{all}$ and a set of candidate entities for each mention}
\KwOutput{Text $T_{i}$ with extracted mentions $\mathbb{M}_{all}$ and a unique disambiguated entity for each mention}
\tcp{First pass - Disambiguate the easy mentions}
$\mathbb{M}_{easy}$ $\leftarrow$ Easy mentions obtained from the first pass on $T_{i}$\;
    $\mathbb{E}_{easy}$ $\leftarrow$ Disambiguated easy entities obtained from the first pass on $T_{i}$\;
    Document Context $C(T_{i})$  $\leftarrow$ $C(T_{i})$ + $\mathbb{E}_{easy}$ \; 
    $\mathbb{M}_{hard}$ $\leftarrow$ $\mathbb{M}_{all}$ - $\mathbb{M}_{easy}$\;
    \tcp{Second pass - Iterate over the hard mentions}
    \ForEach{Mention \emph{$M_{j}$} $\in$ $\mathbb{M}_{hard}$} {
    $\mathbb{H}_{j}$ $\leftarrow$ Candidate entities of $M_{j}$\;
    \tcp{Iterate over the candidate entities of a hard mention}
    \ForEach{Entity \emph{$E_{k}$} $\in$ $\mathbb{H}_{j}$} {
    Entity Context  C'($E_{k}$) $\leftarrow$ C'($E_{k}$) + $\mathbb{E}'_{k}$ (set of easy entities in a window around $E_{k}$) \;
	$F_{E_{k}}$ $\leftarrow$ Generate feature vector of context independent and dependent features values for $E_{k}$ using C'($E_{k}$)\;
	Classify $F_{E_{k}}$ as one of \emph{$\left\{\text{True, False}\right\}$} using Decision Tree classifier\;
	$S_{E_{k}}$ $\leftarrow$ Final score for $E_{k}$ generated using Logistic Regression model weights\;
	Add $S_{E_{k}}$ to set $S_{j}$ (Set of candidate entity scores for \emph{$M_{j}$})\;
    }
    \tcp{Final disambiguation - select one of the candidate entities as disambiguated entity $D_{j}$ for \emph{$M_{j}$}}
    	\uIf{Only one $E_{k}$ $\in$ $\mathbb{H}_{j}$ labeled as  \emph{True}} {
	 	$D_{j}$ $\leftarrow$ $E_{k}$ labeled as \emph{True}\;
	}  \uElse {
	\uIf{Multiple $E_{k}$ labeled as True}{
		$D_{j}$ $\leftarrow$ Highest scoring $E_{k}$ labeled as \emph{True}\;
	}  \uElseIf{None of $E_{k}$ labeled as True}{
		$D_{j}$ $\leftarrow$ $\argmax$ ($S_{E_{k}}$)\;
	}
	\uIf{$D_{j}$ is \textbf{NIL} and \emph{NIL\_MARGIN\_GAIN} < threshold} {
		$D_{j}$ $\leftarrow$ $\argmax$ ($S_{j}$ - $S_{D_{j}}$)\;
	}
}
}
    \Return{Text with extracted mentions and disambiguated entities\;}
      \caption{Lithium EDL algorithm}
    \label{alg:nedl}
\end{algorithm*}


Algorithm \ref{alg:nedl} describes the Lithium EDL two-pass algorithm. 
We discuss it in detail now (the design choices for various parameters are explained in Section \ref{section:tuning}).

\subsection{First pass \label{firstpass}}
The first pass of the algorithm iterates over all mentions in the document text and disambiguates mentions that have:
\begin{itemize}[noitemsep,nolistsep]
\item Only one candidate entity: In this case, the algorithm disambiguates the mention to the lone candidate entity.
\item Two candidate entities with one being \textbf{NIL}/\textbf{MISC}:  In this case, the algorithm disambiguates the mention to the candidate entity with high \emph{Mention-Entity-Cooccurr} prior probability (above $\lambda_{1}$ - Easy Mention Disambiguation threshold with \textbf{NIL}).
\item Three or more candidate entities with one entity mapping with very high prior: In this case, the algorithm disambiguates the mention to the candidate entity with high \emph{Mention-Entity-Cooccurr} prior probability (above $\lambda_{2}$ - Easy Mention Disambiguation threshold). 
\end{itemize}

Mentions disambiguated in the first pass constitute the set $\mathbb{M}_{easy}$ and their corresponding disambiguated entities constitute the set $\mathbb{E}_{easy}$. The remaining ambiguous mentions constitute the set $\mathbb{M}_{hard}$ and are disambiguated in the second pass.

\subsection{Second pass}
The second pass of the algorithm uses several context-independent and context-dependent features as well as supervised classifiers to label and score the candidate entities for each hard mention and finally disambiguate it. 

\subsubsection{\textbf{Features}} \label{sec:features}

We use several language agnostic features to classify each candidate entity for each hard mention as `True' or `False'. These include both context-independent (useful for disambiguating and linking entities in short and sparse texts such as tweets) as well as context-dependent features (useful for disambiguating and linking entities in long and rich text). Each feature produces a real value in [0.0,1.0].

The context independent features are:

\begin{itemize}[noitemsep,nolistsep]

\item \textbf{Mention-Entity Cooccurrence} (\emph{Mention-Entity-Cooccurr}) - This feature value is equal to the \emph{Mention-Entity-Cooccurr} prior probability.

\item \textbf{Mention-Entity Jaccard Similarity} (\emph{Mention-Entity-Jaccard}) - This reflects the similarity between the mention $M_{i}$ and the representative name of a candidate entity $E_{j}$. The mention and the entity display names are first tokenized and the Jaccard similarity is then computed between the token sets as
\small
$$
{{Tokens(M_{i}) \cap Tokens(E_{j})} \over {Tokens(M_{i}) \cup Tokens(E_{j})}}
$$
\normalsize
For instance, the mention \emph{Marvel} could refer to the entities \emph{Marvel Comics} or \emph{Marvel Entertainment}, both of which have a Jaccard Similarity of 0.5 with the mention.

\item \textbf{Entity Importance} (\emph{Entity-Importance}) - This reflects the importance or the relevance of the candidate entity as determined by an entity scoring and ranking algorithm \cite{Bhattacharyya-importance} which ranks the top 1 million entities occurring in our KB. For instance, the entity \emph{Apple Inc.} has an importance of 0.66 while \emph{Apple (fruit)} has an importance of 0.64 as ranked by the Entity Scoring algorithm.

\end{itemize}

\begin{figure}[t]
\centering
\includegraphics[width=75mm ,height=35mm]{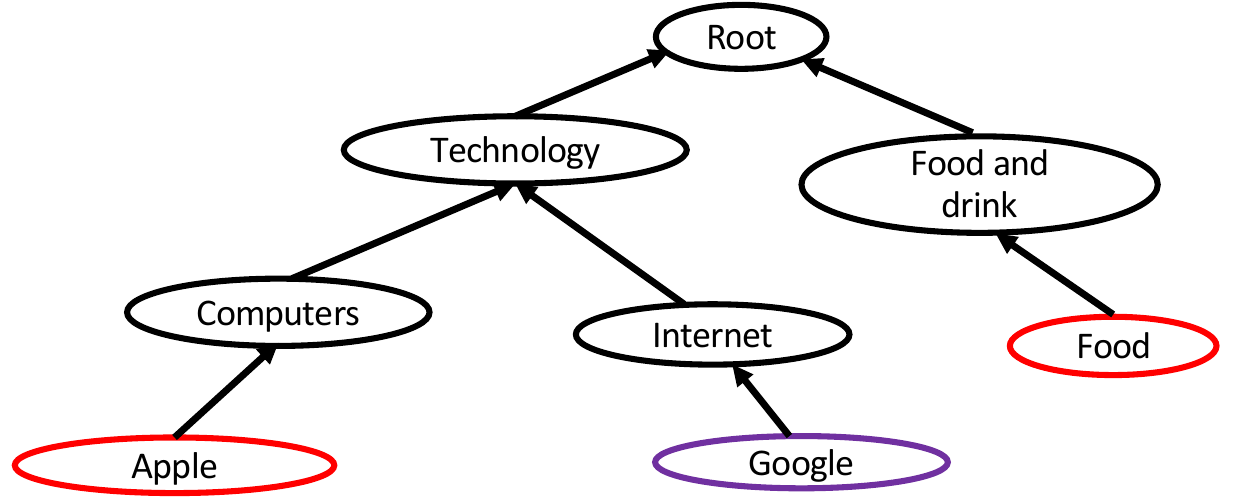}
\caption{Semantic distance between topics in Klout Topic Ontology Space}
\label{fig:topicsimdist}
\end{figure}

For the following context dependent features, we assume that for a candidate entity $E_{i}$, we maintain an entity context C'($E_{i}$) which contains a window $\mathbb{E}'_{i}$ of \emph{W} disambiguated easy entities immediately surrounding $E_{i}$.

\begin{itemize}[noitemsep,nolistsep]
\item \textbf{Entity Entity Cooccurrence} (\emph{Entity-Entity-Cooccurr}) - 
This feature value is equal to the averaged co-occurrence of a candidate entity with the disambiguated easy entities in $\mathbb{E}'_{i}$ and is computed as:
\small
$$
{\sum_{j=1}^{W} Co-occurrence-count (E_{i}, E_{j}) \over {W}} \forall E_{j} \in  \mathbb{E}'_{i}
$$
\normalsize

\item \textbf{Entity Entity Topic Semantic Similarity} (\emph{Entity-Entity-Topic-Sim}) - As mentioned in Section \ref{subsection:OfflineDictionary}, each entity in our KB is associated with a finite number of topics in our topic ontology. For instance, entity \emph{Apple Inc.} maps to the topic `Apple' and \emph{Google Inc.} maps to the topic `Google' while `\emph{Apple (fruit)}' will map to the topic `Food'. Figure \ref{fig:topicsimdist} shows a partial view of the ontology for the above mentioned topics.  

For each candidate entity $E_{i}$ of a hard mention $M_{i}$, we compute the minimum \emph{semantic distance} of its topics with topics of each entity in $\mathbb{E}'_{i}$ over all possible paths in our topic ontology space. The similarity is the inverse of the distance. For instance, consider the hard mention \emph{Apple}, having two candidate entities - \emph{Apple Inc.} and \emph{Apple (fruit)} for it, and $\mathbb{E}'_{i}$ containing the entity \emph{Google Inc.} which has been disambiguated. As shown in Figure \ref{fig:topicsimdist}, the semantic distance between the topics for \emph{Apple Inc.} and \emph{Google Inc.} is 4 while the semantic distance between the topics for \emph{Apple (fruit)} and \emph{Google Inc.} is 5. As a result, it is more likely that \emph{Apple} disambiguates to \emph{Apple Inc}. 

Thus, we first determine the set of topics $\mathbb{T}_{i}$ that the candidate entity $E_{i}$ is associated with. For each entity $E_{j}$ in $\mathbb{E}'_{i}$, we generate the set of topics $\mathbb{T}_{j}$. The feature value is computed as
\small
$$
\max {1 \over distance (t_{i}, t_{j})} \forall t_{i} \in \mathbb{T}_{i}, t_{j} \in \mathbb{T}_{j}
$$
\normalsize
\end{itemize}

\begin{figure*}[t]
\centering
\includegraphics[width=\textwidth,height=80mm]{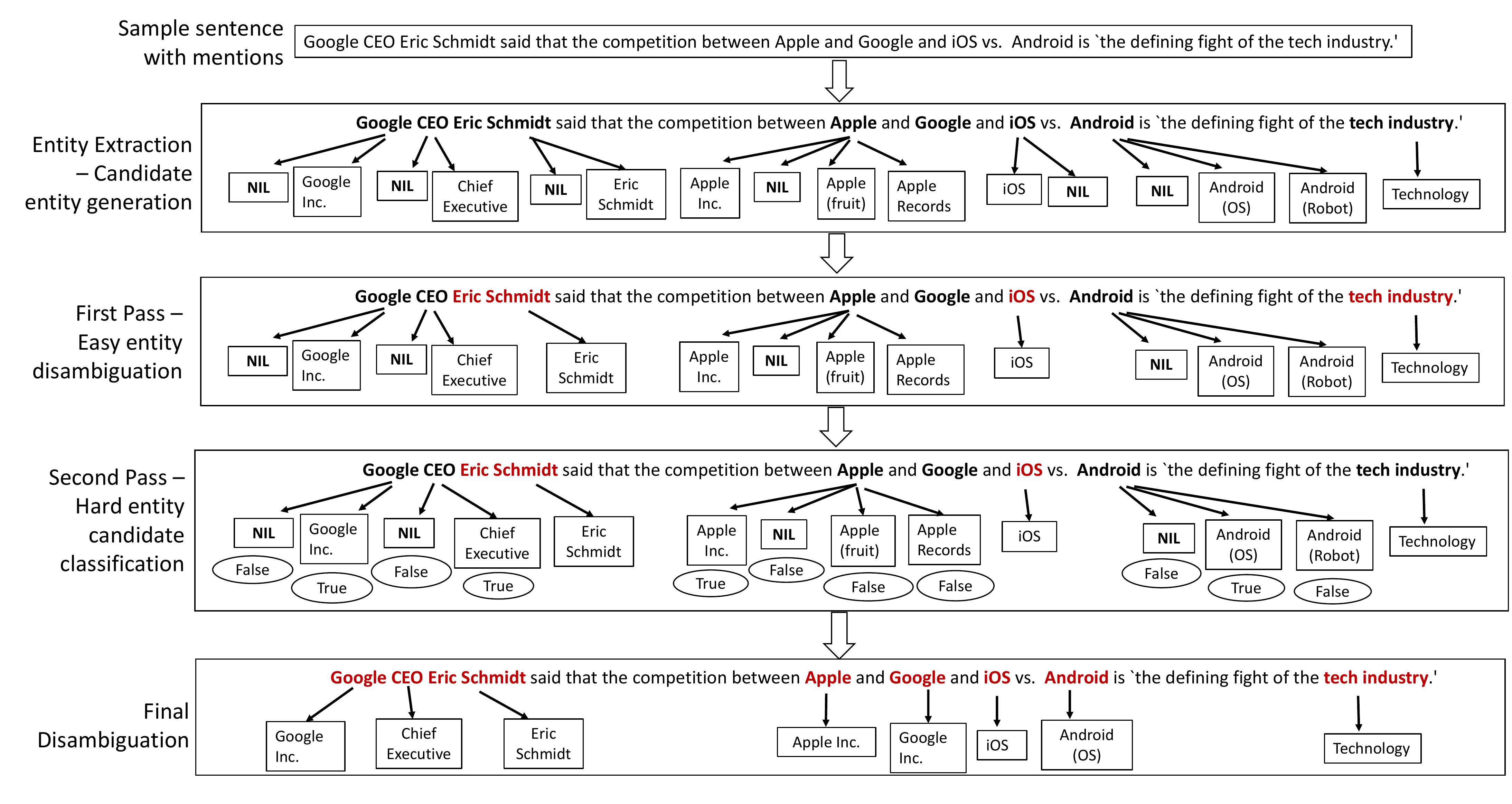}
\caption{Disambiguation of a sample sentence (best viewed in color)}
\label{fig:sample}
\end{figure*}

\subsubsection{\textbf{Classification and Scoring}}

As a penultimate step in the second pass, the computed features are combined into a feature vector for a candidate entity and the Decision Tree classifier labels the feature vector as `True' or `False'. In addition, for each candidate entity, we also generate final scores using weights generated by the Logistic Regression classifier that we trained in Section \ref{subsubsec:class}. We use an ensemble of the two classifiers in the final disambiguation step as it helps overcome the individual bias of each classifier.

\subsubsection{\textbf{Final Disambiguation}}
The final disambiguation step needs to select one of the labeled candidate entities as the disambiguated entity for the mention. However, multiple cases arise at the time of disambiguation:
\begin{itemize}[noitemsep,nolistsep]
\item Only one candidate entity is labeled as `True'- Here, the algorithm selects that entity as the disambiguated entity for the given mention.
\item Multiple candidate entities labeled  as `True'  - Here, the algorithm selects the highest scoring entity (from among those labeled `True') as the disambiguated entity except when this entity is \textbf{NIL}/\textbf{MISC}. In that case, the algorithm checks the \emph{margin of gain} or the score difference between the \textbf{NIL}/\textbf{MISC} entity and the next highest scoring entity that is labeled `True'. If the margin of gain is less than a threshold (less than \textbf{NIL} margin of gain threshold, $\lambda_{3}$) then the next highest scoring entity (from among those labeled `True') is selected.
\item All candidate entities labeled as `False' - Here, the algorithm selects the highest scoring entity as the disambiguated entity except when this entity is \textbf{NIL}/\textbf{MISC}. In that case, the algorithm checks the margin of gain for this entity over the next highest scoring entity. If the margin of gain is less than a threshold (less than \textbf{NIL} margin of gain threshold, $\lambda_{3}$) then the next highest scoring entity is selected.
\end{itemize}

\subsection{Demonstrative Example}\label{sec:demoex}

To demonstrate the efficacy of our algorithm, let's disambiguate the sample text: 
\emph{``Google CEO Eric Schmidt said that the competition between Apple and Google and iOS vs.  Android is `the defining fight of the tech industry.' ".} 

Figure \ref{fig:sample} walks through the disambiguation of the sample text. The Text Preprocessing stages extract the mentions (highlighted in bold) and generate the candidate entities and the prior cooccurrence scores for each mention\footnote{Though our algorithm utilizes the Freebase machine id for each candidate entity, we only show the entity name for clarity.}. 
As shown, the extracted mentions and their candidate entities are:
\begin{itemize}[noitemsep,nolistsep]
\item \emph{Google} - \textbf{NIL} and \emph{Google Inc.}
\item \emph{CEO} - \textbf{NIL} and \emph{Chief Executive}
\item \emph{Eric Schmidt} - \textbf{NIL} and \emph{Eric Schmidt}
\item \emph{Apple} - \textbf{NIL}, \emph{Apple (fruit)}, \emph{Apple Inc.} and \emph{Apple Records}
\item \emph{iOS} - \textbf{NIL} and \emph{iOS}
\item \emph{Android} - \textbf{NIL}, \emph{Android (OS)} and \emph{Android(robot)}
\item \emph{tech industry} - \emph{Technology}
\end{itemize}
In the first pass, the algorithm disambiguates the easy mentions. Based on their high prior scores and number of candidate entities, it disambiguates \emph{Eric Schmidt}, \emph{iOS} and \emph{tech industry} (highlighted in color) to their correct entities. 
In the second pass, it uses the easy mention window and computes several context dependent and independent features to score and classify the candidate entities of the hard mentions. Note that for the purpose of clarity and simplicity, we are not walking through the feature and final score computation. As shown, for the remaining hard entities, it has classified the candidate entities as `True' or `False'.
In the final disambiguation step, it selects one of the labeled entities as the correct disambiguated entity. In the sample sentence, for all the mentions, only one of the candidate entities is labeled as `True', and hence the algorithm selects that entity  as the disambiguated entity for each mention.


\section{Parameter Tuning}
\label{section:tuning}


Our algorithm uses four different hyperparameters - 2 in the first pass and 2 in the second pass. These are:

\begin{itemize}[noitemsep,nolistsep]
\item Easy mention disambiguation threshold with \textbf{NIL} ($\lambda_{1}$) - This threshold is used to disambiguate easy mentions which have 2 candidate entities and one of them is the \textbf{NIL} entity.
\item Easy mention disambiguation threshold ($\lambda_{2}$) - This threshold is used to disambiguate easy mentions which have 3 or more candidate entities but the mention maps to one of them with a very high prior probability.

\item \textbf{NIL} margin of gain threshold ($\lambda_{3}$) - This threshold is used in the second pass to disambiguate entities when multiple or none of the candidates are labeled `True'.
\item Window size (\emph{W}) - This parameter represents the size of the easy entity window around each hard entity.
\end{itemize}

Using the process described in Section \ref{data_set}, we generated a ground truth validation set of 10 English Wikipedia pages which had a total of 7242 mentions and 23,961 candidate entities. We used parameter sweeping experiments to determine the optimal value of these parameters. We measured the performance (in terms of precision, recall and f-score) of the algorithm on the validation set with different parameter settings and picked the parameter values that had the best performance. 
Based on our experiments, we set the optimal value of $\lambda_{1}$ as 0.75, $\lambda_{2}$ as 0.9, \emph{W} as 400 and $\lambda_{3}$ as 0.5. 

\section{Evaluation}
\label{section:evaluation}
\begin{table}[t]
\centering
\begin{tabular}{|c|c|c|}
\hline
 & \multicolumn{2}{|c|}{\textbf{Ground Truth Label}}  \\
 \cline{2-3}
\textbf{Predicted Label}  & Correct Entity & \textbf{NIL} \\
\hline
  Correct Entity &TP & FP\\
\hline
 Wrong Entity& FP & FP\\
\hline
 \textbf{NIL} & FN & TN\\
\hline
\end{tabular}
\caption{Confusion matrix for our EDL system}
\label{PositivesNegatives}
\end{table}

\begin{table*}[!th]
\centering
\begin{tabular}{|c|c|c|c|c|}
\hline
\textbf{Features} & \textbf{Precision} & \textbf{Recall} & \textbf{F-score} & \textbf{Accuracy}\\
\hline
 \emph{Mention-Entity-Cooccurr} (context independent) & 0.65 & 0.75 & 0.70 & 0.62\\
\hline
\emph{Mention-Entity-Jaccard} (context independent) & 0.47 & 0.93 & 0.48 & 0.63\\
 \hline
\emph{Entity-Importance} (context independent) & 0.50 & 0.90 & 0.50 & 0.65\\
 \hline
\emph{Entity-Entity-Cooccurr} (context dependent) & 0.54 & 0.91 & 0.54 & 0.68\\
 \hline
\emph{Entity-Entity-Topic-Sim} (context dependent) & 0.49 & 0.88 & 0.49 & 0.63\\
 \hline
Combined Context independent features & 0.63 & 0.83 & 0.62 & 0.72\\
  \hline
Combined Context dependent features & 0.52 & 0.92 & 0.52 &0.66\\
\hline
All features & 0.63 & 0.87 & 0.73 & 0.64\\
\hline
\end{tabular}
\caption{Precision, recall, f-score and accuracy for different features and feature sets on our test set (English only)}
\label{FeatureEffectiveness}
\end{table*}

\begin{table}[!th]
\centering
\begin{tabular}{|c|c|c|c|c|}
\hline
\textbf{Language} & \textbf{Precision} & \textbf{Recall} & \textbf{F-score} & \textbf{Accuracy}\\
\hline
English & 0.63 & 0.87 & 0.73 & 0.64\\
\hline
French & 0.59 & 0.86 & 0.70 & 0.6\\
 \hline
German & 0.63 & 0.90 & 0.74 & 0.64\\
 \hline
Spanish & 0.58 & 0.89 & 0.70 & 0.60\\
 \hline
Japanese & 0.73 & 0.88 & 0.80 & 0.74\\
 \hline
 \end{tabular}
\caption{Precision, recall, f-score and accuracy across various languages}
\label{LanguagePerf}
\end{table}


\begin{figure}[t]
  \centering
  \begin{subfigure}[b]{0.4\textwidth}
    \includegraphics[width=\textwidth,height=55mm]{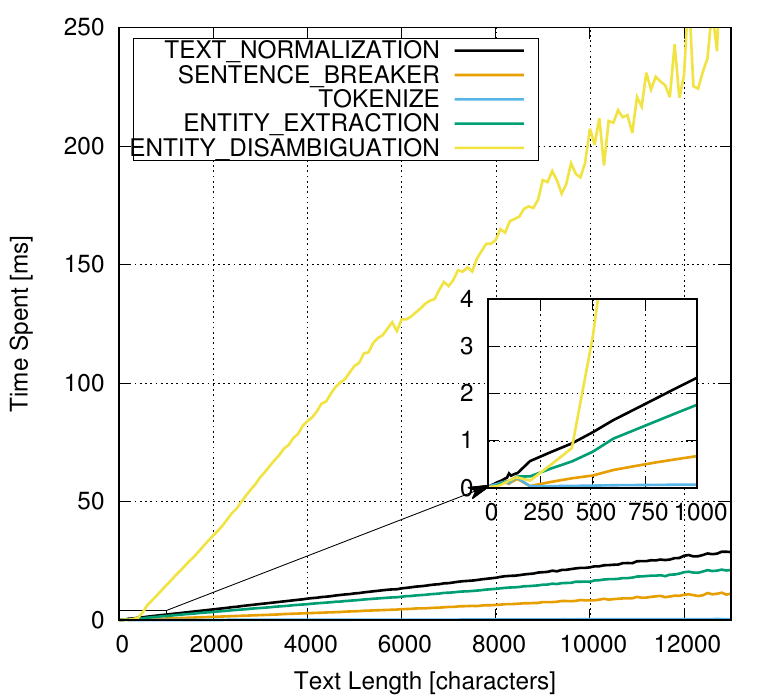}
    \caption{Processing Stages for English}
    \label{fig:processing_times_per_stage}
  \end{subfigure}
  \begin{subfigure}[b]{0.4\textwidth}
    \includegraphics[width=\textwidth,height=55mm]{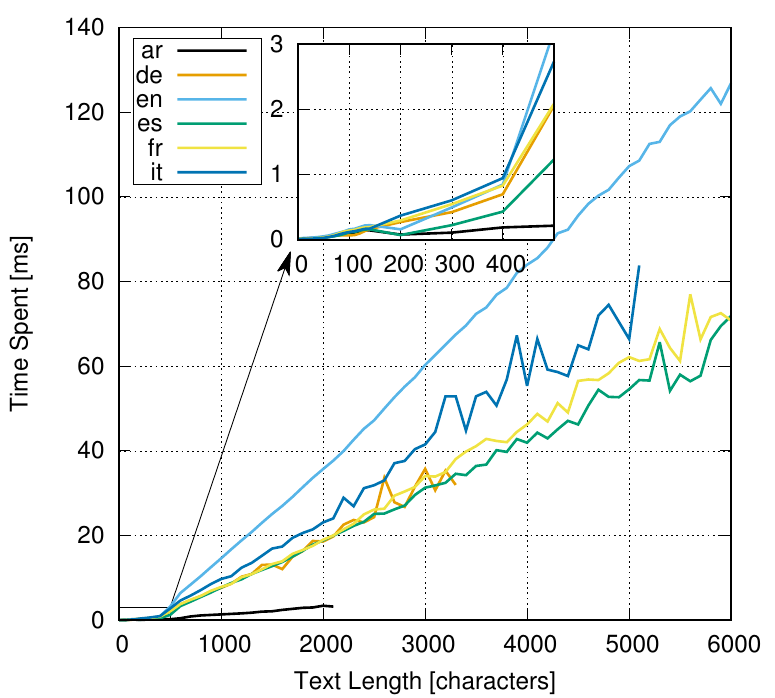}
    \caption{Different Languages for Entity Disambiguation}
    \label{fig:processing_times_per_language}
  \end{subfigure}
  \caption{Processing times as function of text length}
  \label{fig:processing_times}
\end{figure}

\subsection{Test data}
Using the process described in Section \ref{data_set}, we generated a ground truth test set of 20 English Wikipedia pages which had a total of 18,773 mentions. 

\subsection{Metrics}
We use standard performance metrics like precision, recall, f-score and accuracy to evaluate our EDL system on the test set. However, due to our problem setup, we calculate true positives, false positives, and true negatives and false negatives in an unconventional way as shown in Table \ref{PositivesNegatives}. 
Precision, recall, f-score and accuracy are calculated in the standard format as:
 $
 \textnormal{P} = {{\textnormal{$t_{p}$}} \over {\textnormal{$t_{p} $+ $f_{p}$}}}
$,
$
 \textnormal{R}= {{\textnormal{$t_{p}$}} \over {\textnormal{$t_{p}$ + $f_{n}$}}}
$,
$
 \textnormal{F1}= {{2 \times \textnormal{P} \times \textnormal{R}} \over {\textnormal{P + R}}}
$
and 
$
 \textnormal{Accuracy} = {{\textnormal{$t_{p}$ + $t_{n} $}} \over {\textnormal{$t_{p}$ + $t_{n}$ + $f_{p}$ + $f_{n}$}}}
$

\subsection{Results}\label{sec:results}
We compute the performance metrics for individual features as well as for various feature sets on our English language test set to assess their impact. 
Table \ref{FeatureEffectiveness} shows the feature effectiveness results for our algorithm. As evident from the results, \emph{Mention-Entity-Cooccurr} has the biggest impact on the performance of the algorithm among all individual features as it has the highest individual precision and f-score. 

When combined, the context independent features combined 
have higher precision and f-score than the context dependent features. 
This could be due to the fact that in shorter text documents, there may not be enough easy mentions disambiguated in the first pass. 
Since the context dependent features rely on the easy entity window for computation, their performance will be impacted. However, when all these features are taken together, the overall performance improves even further. This demonstrates that context is an important factor in entity disambiguation and linking.
Our final algorithm, which utilizes all the context dependent and independent feature sets, has a precision of 63\%, recall of 87\% and f-score of 73\%. 

Table \ref{LanguagePerf} shows the performance of the Lithium EDL system across various languages. We note that the test datasets for these languages are smaller. However, the algorithm's performance is comparable to that for the English dataset.


\subsection{Runtime Performance}

The Lithium EDL system has been built to run in a bulk manner as well as a REST API service.
The two major challenges that we faced while developing the system were the volume of new data that we process in bulk daily and limited computational capacity. These challenges had a significant influence on our system design and algorithmic approach.

As a demonstrative example, the most consuming task in our MapReduce cluster processes around 910 million documents, with an average document size of 169 bytes, taking about 2.2ms per document. Our MapReduce cluster has around 150 Nodes each having a 2.5 GHz Xeon processor. The processing is distributed across 400 reducers. The Reduce step takes about 2.5 hrs. Each reducer task runs as a single thread with an upper bound of 7GB on memory where the processing pipeline and models utilize 3.7GB. 

A more detailed breakdown of the computational performance of our system as a function of document length is shown in Figure \ref{fig:processing_times}. The overall performance of the system is a linear function of text length. We also analyze this performance for different languages as well as for different stages of the Lithium NLP pipeline. We can see that the computation is slowest for English since it has the maximum number of entities \cite{Spasojevic:dawt}.


\section{Comparison with other commercial systems}
\label{section:relatedwork}

\begin{table}[t]
\begin{minipage}[b]{1\linewidth} 
\centering
\begin{tabular}{|c|c|c|c|}
\hline
 & \textbf{Lithium} & \textbf{Google NL} & \textbf{Both} \\
\hline
  \textbf{English} & 5548 & 1501 & 1062 \\
\hline
  \textbf{Spanish} & 2410	 & 1152 & 839 \\
\hline
  \textbf{Japanese} & 1631 & 801 & 549 \\
\hline
  \textbf{All} & 9589 & 3454 & 2450 \\
\hline   
\end{tabular}
\caption{Comparison of Lithium EDL and Google Cloud NL API}
\label{table:ComparisonWithGoogle}
\end{minipage}

\begin{minipage}[b]{1\linewidth}
\centering
\begin{tabular}{|c|c|c|}
\hline
 & \textbf{Lithium} & \textbf{OpenCalais}  \\
\hline
  \textbf{English} & 5548 & 1295 \\
\hline
  \textbf{Spanish} & 2410 & 885 \\
\hline
  \textbf{French} & 3341 & 1161 \\
\hline
  \textbf{All} & 11299 & 3341 \\
\hline   
\end{tabular}
\caption{Comparison of Lithium EDL and OpenCalais API}
\label{table:ComparisonWithOpenCalais}
\end{minipage} 
\end{table}

\begin{figure}[t]
  \centering
  \includegraphics[width=0.47\textwidth,height=70mm]{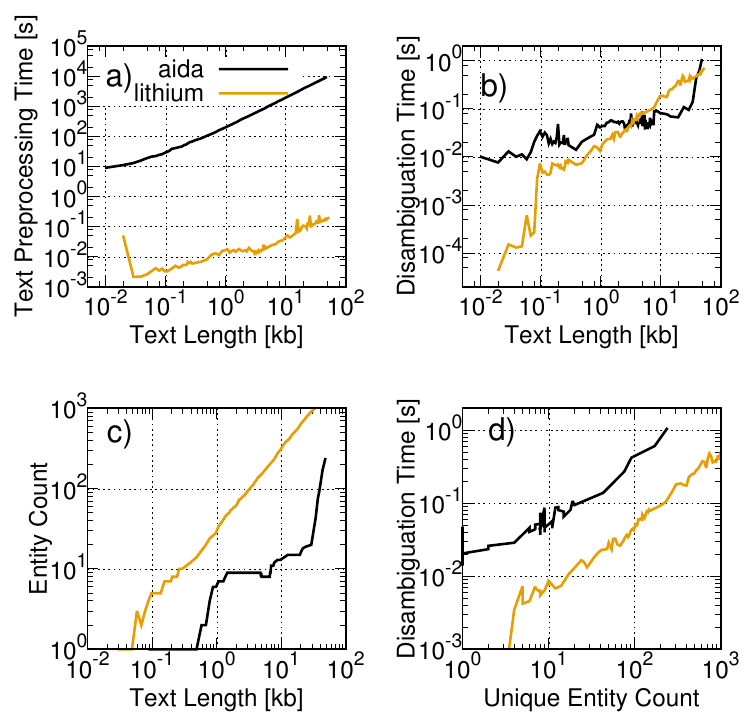}
  \caption{AIDA and Lithium NLP Pipeline Comparisons. \textbf{a)} Text preprocessing runtime;
  \textbf{b)} Disambiguation runtime; \textbf{c)} Extracted entity count;
  \textbf{d)} Disambiguation runtime as function of entity count;}
  \label{fig:aida_vs_lithium}
\end{figure}

\begin{figure*}
\centering
\includegraphics[width=\textwidth]{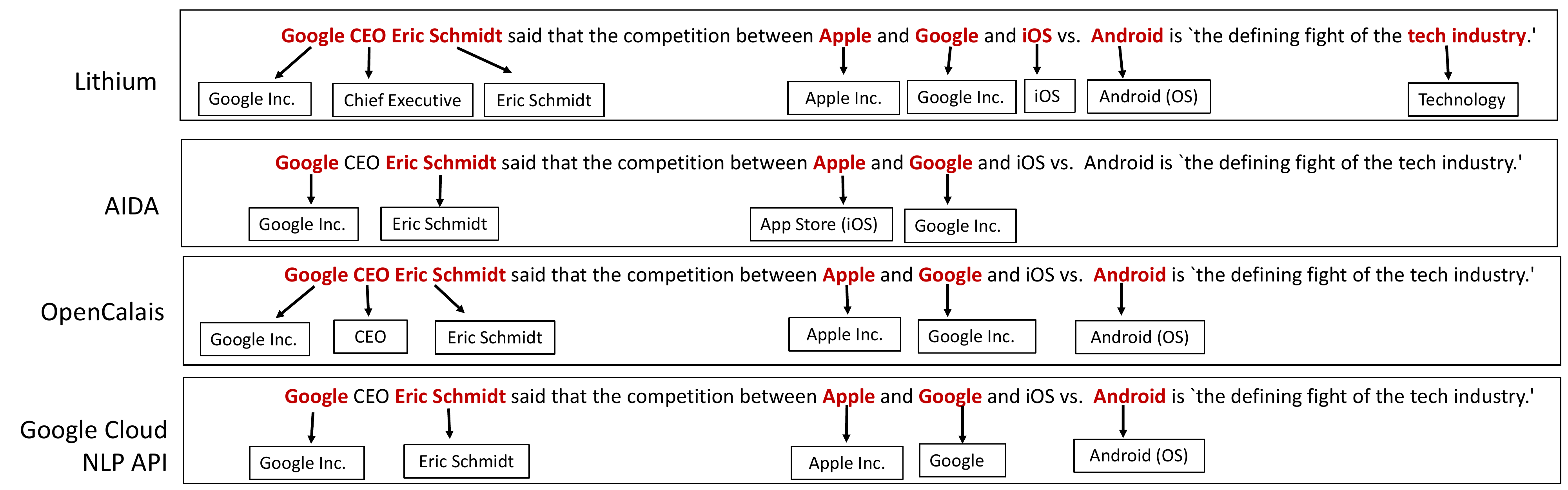}
\caption{Comparison of the different systems on our demonstrative example}
\label{fig:samplecomparison}
\end{figure*}

Currently, due to limited resources at our end and due to inherent differences in KB, data and text preprocessing stages, a direct comparison of the Lithium EDL system's performance (in terms of precision, recall and f-score) with other commercial systems, such as Google Cloud NL API, OpenCalais and AIDA, is not possible. Hence, we compare our system with them on a different set of metrics.

\subsection{Comparison on languages}
While the Lithium EDL system supports about 6 different languages (English, Arabic, Spanish, French, German, Japanese), Google Cloud NL API supports mainly 3 languages: English, Spanish, and Japanese. Similarly, OpenCalais supports only English, Spanish, and French while AIDA only supports English and Arabic.

\subsection{Comparison on linked entity density} \label{sec:entitydensitycomparison}

A major advantage of our system is the ability to discover and disambiguate a much larger number of entities compared to other state-of-the-art systems.
As a demonstration, we compared our result with Google Cloud NL API and OpenCalais\footnote{We also analyzed AlchemyAPI (\url{http://www.alchemyapi.com/resources}) but it only processed a limited amount of text in a document and was not very stable on languages other than English.}.
In particular, we ran both APIs on documents in our test data set with the common subset of languages that they supported.

Table \ref{table:ComparisonWithGoogle} compares the total number of unique entities disambiguated by Lithium EDL system and those by Google NL. 
An entity from Google NL is considered to be disambiguated if it was associated with a Wikipedia link.
Column \textbf{Both} shows the numbers of entities that were disambiguated by both systems. 
Most entities disambiguated by 
Google NL were also disambiguated by our system. In addition, our system disambiguated several more entities. Based on the the precision of our system, we can estimate that at least 6080 disambiguated entities from our system are correct. This implies that Google NL missed more than 2600 entities that were correctly disambiguated by our system. Thus, our system correctly disambiguated at least 75\% more entities than Google NL. 

Table \ref{table:ComparisonWithOpenCalais} shows a similar comparison between our system and OpenCalais.
Every entity from OpenCalais API is considered to be disambiguated. 
However, since OpenCalais entity does not link the disambiguated entities to Wikipedia or Freebase but to their own proprietary KB, we cannot determine which entities were discovered by both the systems. 
Nevertheless, based on the precision of our system, at least 3500 entities that were correctly disambiguated by our system, were missed by OpenCalais, which is significantly more than the number of entities they detected.

\subsection{Comparison on runtime}

We compared the runtime performance of the Lithium NLP pipeline against AIDA\footnote{\url{https://github.com/yago-naga/aida}} \cite{nguyen2014aida} on several English language documents. Comparison results are shown in Figure \ref{fig:aida_vs_lithium} on the log-log scale.
In Figure \ref{fig:aida_vs_lithium}a we can see that the text preprocessing stage of the Lithium pipeline is about 30,000-50,000 times faster compared to AIDA which is based on Stanford NLP NER \cite{Finkel:2005:INI:1219840.1219885}.
The results for the disambiguation stage are shown in Figure \ref{fig:aida_vs_lithium}b. The disambiguation stage for both the systems take a similar amount of time.
However, AIDA fails to extract as many entities as evident in Figure \ref{fig:aida_vs_lithium}c which shows that AIDA extracts $2.8$ times fewer entities per $50kb$ of text. Finally, the disambiguation runtime per unique entity extracted of Lithium pipeline is about $3.5$ times faster than AIDA as shown in Figure \ref{fig:aida_vs_lithium}d. In conclusion, although
AIDA entity disambiguation is fairly fast and robust, our system's runs significantly faster and is capable of extracting many more entities.

\subsection{Comparison on demonstrative example}
In order to explicitly demonstrate the benefits and expressiveness of our system, we also compare the results of our EDL system with Google Cloud NL API, OpenCalais and AIDA on the example that we discussed in Section \ref{sec:demoex}. Figure \ref{fig:samplecomparison} shows the disambiguation and linking results generated by our EDL system and the three other systems (Google NL Cloud API, OpenCalais and AIDA) that we compare with. As evident, our EDL system disambiguates and links more entities correctly than the other 3 systems. All the other systems fail to disambiguate and link \emph{iOS} and \emph{tech industry}. In addition, AIDA incorrectly disambiguates \emph{Apple}.

\section{Conclusion and Future Work}
\label{section:conclusion}
In this paper, we presented the Lithium EDL system that disambiguates and links entity mentions in text to their unique Freebase ids. Our EDL algorithm uses several context dependent and context independent features to disambiguate mentions to their respective entities. Moreover, it recognizes several types of entities in addition to named entities like people, places, organizations. 
In addition, our EDL system is language-agnostic and currently supports several languages including English, Arabic, Spanish, French, German, and Japanese. As a result, it is highly applicable to process real world text such as multi-lingual user generated content from social media in order to model user interests and expertise. 

We compared our EDL system with several state-of-the-art systems and demonstrate that it has high throughput and is very lightweight. It can be run on an off-the-shelf commodity machine and scales easily to large datasets. Also, our experiments show that our EDL system extracts and correctly disambiguates about 75\% more entities than existing state-of-the-art commercial systems such as Google NLP Cloud API and Open Calais and is significantly faster than some of them.
In future, we plan to add support for several other languages to our EDL system once we have collected enough ground truth data for them. We also plan to migrate to Wikipedia as our KB. We will also compare our system's performance against several state-of-the-art systems on metrics such as precision, recall and f-score with respect to existing benchmarked datasets.

\bibliographystyle{abbrv}
\bibliography{bibliography}

\end{document}